\documentclass[twocolumn,english,aps,prb,floatfix,preprintnumbers,showpacs,amsfonts,amssymb,superscriptaddress]{revtex4-1}

\usepackage{amsmath}
\usepackage{graphicx}
\usepackage{dcolumn}
\newcommand{\mypath}[1]{./#1}

\begin{document}

\title{Complex state found in the colossal magnetoresistance regime of \\
models for manganites}

\author{Cengiz \c{S}en}

\author{Shuhua Liang}

\author{Elbio Dagotto}
\affiliation{Department of Physics and Astronomy, The University of Tennessee, Knoxville, TN 37996}
\affiliation{Materials Science and Technology Division, Oak Ridge National Laboratory, Oak Ridge, TN 32831}

\date{\today}

\begin{abstract}
The colossal magnetoresistance (CMR) effect 
of manganites is widely believed to be caused by the
competition between a ferromagnetic (FM) metallic state 
induced by the double-exchange mechanism
and an insulator with complex spin, charge, and orbital order. 
Recent computational studies in small clusters
have indeed reported a CMR precisely near 
the frontier between those two states
at a realistic hole density $x=1/4$.
However, the detailed characteristics of the competing insulator were not fully understood 
in those previous investigations.
This insulator is expected to display special properties that lead to the CMR, 
otherwise any competition
between ferromagnetic and antiferromagnetic states would induce such an effect, 
which is not the case experimentally. 
In this report, the competing insulator at electronic density $x=1/4$ and in 
the CMR regime is studied in detail using the double-exchange two-orbital model  with 
Jahn-Teller lattice distortions on two-dimensional clusters, 
employing a careful large-scale 
cooling down process in the Monte Carlo simulations to avoid being trapped in metastable states. 
Our investigations show that 
this competing insulator has an unexpected complex structure,
involving diagonal stripes with alternating regions displaying FM and CE-like order. The
level of complexity of this new state 
even surpasses that of the recently unveiled spin-orthogonal-stripe
states and their associated high degeneracy.
This new state complements the long-standing scenario
of phase separation, since the alternating FM-CE pattern appears even 
in the clean limit. The present and recent investigations are also in
agreement with the many ``glassy'' characteristics of the CMR state found experimentally,
due to the high degeneracy of the insulating states involved in the process.
Results for the spin-structure factor of  
the new states are also here provided to facilitate
the analysis of neutron scattering experiments for these materials.
\end{abstract}


\maketitle

\section{Introduction}

The colossal magnetoresistance (CMR) effect of the manganites 
provides a fascinating example of the collective behavior and unexpected nonlinearities that can
emerge in complex materials, such as transition metal oxides, when several degrees of 
freedom are simultaneously active.\cite{otherreviews,review}
Reaching a complete theoretical understanding of the CMR effect 
is certainly important in the context of Mn oxides, but also
to provide a paradigm for rationalizing related complex phenomena,
or to predict similar effects in other materials.\cite{science}
The current widely accepted scenario to understand the CMR relies on
the existence of insulating states competing with the ferromagnetic (FM) metallic state
induced by the well-known double-exchange mechanism.\cite{review} In addition, particularly
in the presence of small amounts of quenched disorder or other effects such as strain, 
intriguing nanometer-scale electronic 
structures have been found in both experimental efforts
and theoretical studies.\cite{otherreviews,review,science,burgy,verges,kumar,sen06,sen06-fragil,sen06-polynomial,PG,lynn,argyriou,mathieu,bicritical,billinge,aeppli,fye09} 

Considerable progress in the theoretical analysis 
of the CMR effect was recently reached when the CMR effect was numerically found
in a standard two-orbital double-exchange model for manganites at $x=1/4$ doping, including Jahn-Teller
distortions, in the clean limit.\cite{sen07,sen10} The absence of impurities in this study brings the
need to fine tune couplings, such as the antiferromagnetic (AFM) 
superexchange among the
$t_{2g}$ electrons, to locate the system very 
close to the transition from the metal to the insulator
at low temperatures. Quenched disorder is known to enhance considerably the CMR effect and avoid
the fine tuning of couplings, rendering the presence of the effect more universal. However,
it is still an important conceptual issue to study the CMR in the
clean limit using finite clusters, 
even if a fine-tuning of couplings is needed. Such studies provide insight
on what makes the manganite oxides special, since 
few other complex materials present such
a large magnetoresistance effect in so many members of the same oxide family and in
wide ranges of electronic composition and bandwidth. Moreover, a deep understanding
of the CMR effect would contribute 
to new areas of research where manganites play a fundamental
role, such as oxide heterostructures\cite{hetero} and manganite multiferroics.\cite{multiferroics}

There have been two interesting recent conceptual developments in the study of models
for manganites in the clean limit  that are of relevance for the present effort. 
As already briefly mentioned, in Ref.~\onlinecite{sen10} a study
of the conductance vs. temperature, varying the $t_{2g}$ superexchange coupling
to interpolate between the FM metal and the AFM insulator at $x=1/4$, unveiled
a clear CMR effect and a first-order magnetic transition that is 
in good qualitative agreement with experiments.
The AFM insulator was characterized as a C$_{1/4}$E$_{3/4}$ 
state in Ref.~\onlinecite{sen10} due
to the existence in the static spin structure factor 
of a peak at momenta ($\pi$/2,$\pi$/2) which is often taken as evidence
of CE-like states in experimental investigations. Since the dominant state at $x=1/2$ is 
precisely a well-known CE state,\cite{review} then such an assignment was natural, but
this assumption will be revisited below.
The second, even more recent, development was presented in Refs.~\onlinecite{shuhua,SOS}
where surprisingly highly-degenerate diagonal ``stripe'' phases were reported 
at special hole densities in the
range between 0 and 1/2. These investigations showed that the competing 
insulators to the FM metal states at intermediate densities are far more complex than previously
anticipated, particularly in the range of hole doping which is of the 
main interest for CMR effects. 

In other words, in the limit $x=1/2$ it has been clearly shown
both experimentally and theoretically that robust states of the CE form, displaying 
spin, charge, and orbital order, are very stable and dominant. In the other limit $x=0$, there is
also clear evidence of
an A-type AFM insulator with orbital order that is also very dominant. However,
at intermediate hole densities between $x=0$ and $0.5$ 
it is still not fully clear what kind of competitors 
to the FM state do emerge from the many active degrees of freedom in manganites.
The C$_{1-x}$E$_x$ states\cite{hotta03} are natural candidates, 
as assumed in Ref.~\onlinecite{sen10}, but the recent prediction of a competing 
spin-orthogonal-stripe (SOS) states at small electron-lattice
couplings,\cite{SOS} as well as the high 
degeneracy reported in Ref.~\onlinecite{shuhua} that
was discussed above, are
clearly challenging our understanding of this fundamental aspect of the CMR
effect. Knowing with accuracy the properties of the competing insulator state
in the CMR regime will contribute to unveiling 
why this CMR effect occurs in the first place,
since it appears that simply having a competition between a FM metal and any
AFM insulator is not sufficient to have a CMR effect. The AFM insulator must
have special properties that are currently unknown.

Motivated by these recent developments, in the present publication
the insulator found in the $x=1/4$ CMR regime of Ref.~\onlinecite{sen10}
has been further investigated. To our surprise, this insulating state reveals
a remarkable complexity that rivals that of the SOS states and their
corresponding degenerate states. The CE peak at ($\pi$/2,$\pi$/2) that was reported
in Ref.~\onlinecite{sen10} certainly is confirmed, and in fact the overall
insulator does have regions with the characteristic CE zigzag chains.
However, and this is the main result of our effort, the complete state also contains
FM regions of equal length size, such that the overall 
state is a mosaic of CE and FM regions
regularly spaced. Both these CE and FM regions form diagonal 
stripes, that alternate in a ...-FM-CE-FM-CE-... pattern along 
one diagonal. The SOS states\cite{SOS} are not as complex since they 
have the same pattern of spins along the
diagonal stripes, albeit spin rotated by 90$^{\rm o}$ degrees.
However, the new state reported here mixes two very different states, the CE and FM
states, in a combination that is stable in our computer simulations on finite
clusters even
in the clean limit, namely without the need to have disorder or strain
effects to create such nanoscopic phase competition. It is this exotic
combination of CE and FM regions that triggers the CMR effect in our
computational studies once couplings are tuned to the vicinity of the 
metal-insulator regime. As discussed in the conclusions section of this
manuscript, the present results create
a variety of additional conceptual questions in manganites, and they show that the
level of complexity of these materials, at least within the model Hamiltonian
framework and in two-dimensional geometries, is far larger than anticipated in
previous investigations. 

\section{Model and the Monte Carlo Technique} 

In this manuscript, the standard two-orbital lattice Hamiltonian for manganites 
will be studied using two-dimensional finite clusters. In the well-known limit of an
infinite-Hund coupling, the model is explicitly defined as: 
\begin{widetext}
\begin{eqnarray}
  H &=& -\sum_{{\bf ia}\gamma \gamma'\sigma}
  t^{\bf a}_{\gamma \gamma'} 
  [\cos({\theta_{\bf i}/2})\cos({\theta_{\bf i+a}/2})
  +e^{-i(\phi_{\bf i}-\phi_{\bf i+a})}\sin({\theta_{\bf i}/2})\sin({\theta_{\bf i+a}/2)}]
  d_{{\bf i} \gamma \sigma}^{\dag}  d_{{\bf i+a} \gamma' \sigma}
  + J_{\rm AF} \sum_{\langle {\bf i,j} \rangle}
  {\bf S}_{\bf i} \cdot {\bf S}_{\bf j}  \nonumber \\
  &+& \lambda \sum_{\bf i}
  (Q_{1{\bf i}}\rho_{\bf i} + Q_{2{\bf i}}\tau_{{\rm x}{\bf i}} 
  +Q_{3{\bf i}}\tau_{{\rm z}{\bf i}})
  +(1/2) \sum_{\bf i} (\Gamma Q_{1{\bf i}}^2
  +Q_{2{\bf i}}^2+Q_{3{\bf i}}^2),
\end{eqnarray}
\end{widetext}
where $t^{\bf a}_{\gamma \gamma'}$ is the hopping amplitude for the e$_g$ orbitals $\gamma=x^2-y^2$ and 
$\gamma^{\prime}=3z^2-r^2$ in the ${\bf a}$ direction,
$J_{\rm AF}$ is the antiferromagnetic coupling between t$_{2g}$ spins on neighboring 
sites ${\bf i}$ and ${\bf j}$ of a two-dimensional lattice, and
the ${Q}$'s are the various Jahn-Teller (JT) modes (defined extensively 
in previous literature).\cite{review} 
$\lambda$ is the dimensionless electron-lattice coupling constant, and
$\tau_{{\rm z}{\bf i}}=\sum_{\sigma}(d_{{\bf i} a\sigma}^{\dag}d_{{\bf i}a\sigma}
-d_{{\bf i} b\sigma}^{\dag}d_{{\bf i}b\sigma})$ are the pseudo-spin operators defined in Ref.~\onlinecite{review}.
The last term represents the potential (elastic energy) for the distortions, 
with $\Gamma$ the ratio of spring constants for 
breathing- and JT-modes.  The rest of the notation is standard. 
Throughout the numerical simulations described below, cooperative 
lattice distortions  are used, where the actual displacements
$u_{i}$s for the oxygen atoms are the explicit variables in the 
Monte Carlo sampling, instead of the linear combinations $Q_{i}$s that
make up the individual Jahn-Teller modes. Note that in our study both the $t_{2g}$ spins and the lattice degrees
of freedom are considered classical for simplicity, approximation 
widely employed in previous efforts.\cite{review}

The Monte Carlo technique used in these simulations has
been extensively discussed  in the past,\cite{review}
and details will not be repeated here. The procedure to calculate conductances is also
standard.\cite{verges00}
However, an important improvement that has been employed 
here and in some recent studies
merits a more detailed discussion. In the present effort, 
the cool-down method has been used, rather than a more
standard procedure where a particular temperature is chosen 
and the simulation is run to converge to a 
particular set of equilibrium configurations. 
The latter has proved to be problematic in some cases because
there may be competing meta-stable states where the Monte Carlo states become ``trapped'' 
and this prevents a proper convergence
to the true ground state. The cool-down method, described below, is systematic and successful in achieving a better convergence to the ground state, 
the only problem being the increased CPU time required at the beginning of the process.

The cool-down method starts by selecting a temperature grid. Often a denser temperature grid is chosen at low temperatures
since it is in that range that metastability problems are more likely to arise.
The simulation starts in practice in our study at a very high temperature 
$\beta=3$ with a completely random configuration of the classical spins and oxygen
lattice degrees of freedom. As a first step,
the system is allowed to thermalize for the first $10,000$ MC steps, and measurements are taken during the next $5,000$ steps.
Then, the temperature is lowered to the next one in the temperature grid, 
e.g. $\beta=4$, and the simulation process continues all
the way down to $\beta=300$, with $5,000$ steps for thermalization and another $5,000$ steps for measurements at each
temperature.

In general, the acceptance ratios of the Monte Carlo simulations deteriorate significantly with decreasing temperature. 
To avoid this problem, the acceptance ratios were monitored during the entire simulation to make sure that they do not fall below 
a (less than ideal but) reasonable rate ($\approx \%10$) for the lowest temperature simulated in the calculation, $\beta=300$. 
In practice it was observed that a Monte Carlo window 
of $\Delta = 3/\beta$ is sufficient to guarantee that 
the acceptance ratios are still reasonable (here the Monte Carlo window
denotes the amount by which the oxygen coordinates, and the two angles
needed for each classical spin, are modified before the standard
Monte Carlo procedure is used to accept or reject the new configuration).

For a fixed set of model parameters (the electron-lattice coupling $\lambda$, and the superexchange coupling
$J_{\rm AF}$), the cool-down process alone might still not be sufficient for full convergence. However, many sets
of these  model parameters were used in the Monte Carlo runs in parallel (employing several computer nodes)
and in the end, by mere comparison of energies, it was observed that often for a subset  of those model parameters a convergence 
to the true ground state was found.
The existence of a possible new ground state is then confirmed by comparing its energy with those of the neighboring states in the phase 
diagram. In short, by monitoring the smoothness of the values corresponding to several 
Monte Carlo observables when  the model parameters are slightly modified, the overall 
convergence quality of the results can also be monitored.

Once the true ground state for a fixed model parameter set is identified, 
the process is reversed, and this time a ``heat-up'' procedure is carried out, 
namely the simulation starts by using an initial
configuration which is the properly converged last configuration of the cool-down process at the lowest temperature. 
These extra steps increase the chances
that true converged quantities are obtained. The observables at each temperature are calculated
at this stage to a good accuracy by running the simulation with an initial configuration borrowed from the
last configuration of the heat-up process. At this stage in the simulation, $20,000$ Monte Carlo steps
for thermalization were used, and another $100,000$ steps for measurements, where the measurements are taken at every five Monte Carlo steps.
All these values of the Monte Carlo parameters and convergence process are kept the same for all the results discussed below, unless otherwise 
is indicated explicitly.

\section{Quarter doping, $x=1/4$}

\begin{figure}[ht]
\includegraphics[clip,width=8.5cm,height=6cm]{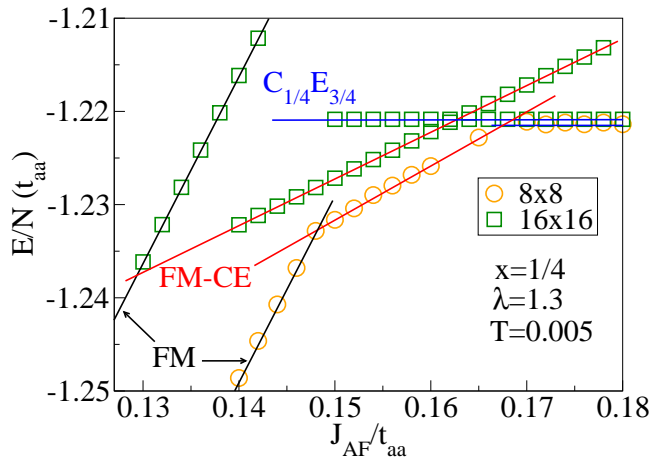}
\caption{(color online) 
Monte Carlo total energy per site vs. $J_{\rm AF}$ for the lowest temperature ($\beta=200$) used in the cool-down process, as
described in the text. Results are obtained using $8\times 8$ and $16\times 16$ lattices. 
Other parameters used are as indicated. Symbols represent the actual MC results.
Error bars are smaller than the size of the symbols (not shown). 
Lines provide guides to the eye.}
\label{Fig_energy_x14}
\end{figure}

In this section, the results obtained using a hole 
doping density $x=1/4$ are presented. In Fig.~\ref{Fig_energy_x14}, 
the Monte Carlo total energy vs. the superexchange coupling $J_{\rm AF}$ is shown at a low temperature $\beta=200$ for
the case of a moderate electron-lattice coupling $\lambda$. In fact, this coupling has been fixed to $\lambda=1.3$ throughout
the $x=1/4$ simulations discussed below, 
since previous investigations have shown that this value is optimal
for the existence of  the CMR peak in the resistivity vs. temperature calculations.\cite{sen10}

\begin{figure}[ht]
\flushleft{
\includegraphics[clip,width=8.6cm,height=6.2cm]{\mypath{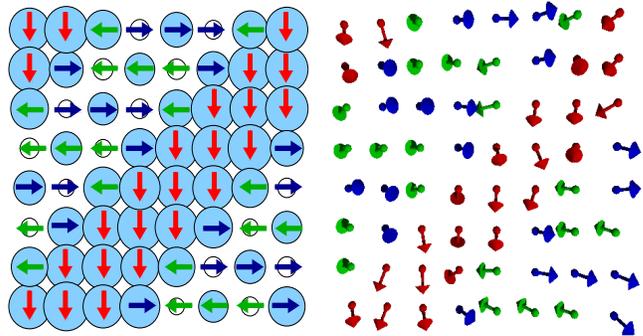}}}
\caption{(color online) 
Real-space representation of the new intermediate state ``FM-CE'' found at $x=1/4$. 
The idealized spin configuration (left) is derived from
the actual Monte Carlo converged state (right) obtained at $\lambda=1.3$ and $J_{\rm AF}=0.16$. 
The local charge densities that are depicted with circles in the idealized configuration 
are taken from the same Monte Carlo snapshot shown on the right. 
Densities that are $\langle n \rangle >0.65$ are shown with light-blue filled 
circles. The largest circle displayed in the figure corresponds to $\langle n \rangle \approx 1$, 
showing that the ferromagnetically
oriented portion of the new state has approximately that density. }
\label{Fig_intx14}
\end{figure}

Three states are here identified as ground states at this particular doping, including the
new state that is the focus of the present publication. 
At small $J_{\rm AF}$, the ground state is the well-known 
FM metal induced by the double-exchange mechanism. 
At large $J_{\rm AF}$ in the range shown in Fig.~\ref{Fig_energy_x14}, 
the ground state is a previously discussed $C_{1/4}E_{3/4}$-type charge-ordered 
insulator.\cite{hotta03} Figure 1 of Ref.~\onlinecite{sen10} shows that this state was
previously believed to be the main competitor to the FM metallic state. 
The main surprise, and main result of this publication,
arises at the intermediate superexchange couplings shown in Fig.~\ref{Fig_energy_x14}. 
In this regime, a new intermediate 
state is identified since the slope of the energy curve is different
from the two extreme cases with purely FM or CE characteristics. 
The intermediate state must actually be a combination of the properties
of the two neighboring states, since it is stable in a narrow region between them.
In fact, this novel state consists of diagonal FM regions located in real space 
sandwiched between CE-like zigzag
spin arrangements of a similar size in length, as shown in Fig.~\ref{Fig_intx14}. 
This new state will be called ``FM-CE'' in the present publication for simplicity.
At low temperatures, the level crossing of the energies of the FM-CE state
with the competitors FM and CE states signals a first-order transition.
While the FM regions do not display any strong indication of charge ordering, 
it is clear from the Monte-Carlo
snapshots that the CE-like regions are charge-ordered with a checkerboard pattern 
(see Fig.~\ref{Fig_intx14}).
Note also that the FM regions present an electronic density 
close to 1 (i.e. $x=0$), namely close to the undoped limit, while
the average density of the CE region is close to $x=1/2$, rendering the overall
density $x=1/4$. Thus, this state presents a spontaneous nanometer-scale charge phase
separation even in the clean limit here investigated. 

It is important to remark that very time consuming simulations on a $16\times 16$ 
lattice have also been carried out to investigate if the new phase is a spurious
consequence of size effects on the $8\times 8$ cluster. The $16\times 16$ simulations
clearly indicate that the new state is present in this lattice,
actually on a {\it wider} range of the $J_{\rm AF}$ coupling, suggesting that the 
bulk system will also display the same novel FM-CE state 
found in the simulations presented here.
Note that for the $16\times 16$ it was not practical to use the same number of MC steps
as used for the $8\times 8$ lattice. Instead, 
$1,000$ thermalization and $2,000$ measurement steps were used
in the former. Also, for the three competing 
states that are considered here, the spins were frozen to the
ideal configurations during the simulation, and the simulations started with the converged
phononic configurations that were borrowed from the $8\times 8$ lattice. The smoothness
of the results at various couplings indicate that a good convergence has been reached in
our study.

\begin{figure}[ht]
\flushleft{
\includegraphics[clip,width=8cm,height=6cm]{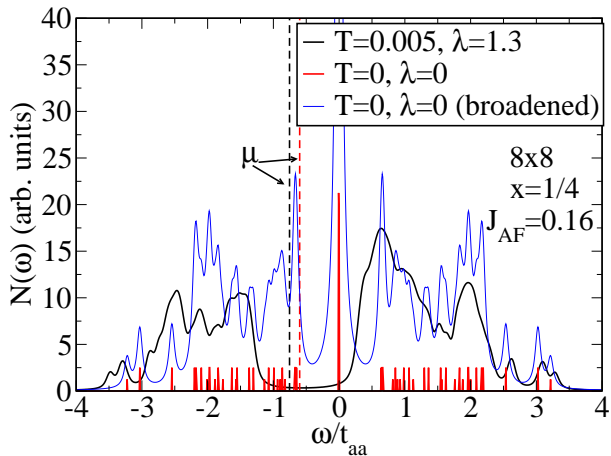}}
\caption{(color online) 
Density-of-states for the new FM-CE state shown for an 
idealized ($T=0$, $\lambda=0$, fixed spins) and a Monte Carlo converged
spin configurations at density $x=1/4$ ($T=0.005$, $\lambda=1.3$). 
The location of the chemical potential is indicated 
with $\mu$ and it is shown with dashed lines. The density-of-states and
the resistivity calculations (see Fig.~\ref{Fig_resistivity2}) 
confirm that the state at $\lambda=1.3$ is insulating. 
The rest of the parameters used are indicated in the figure. Notice
that there are states populated at the Fermi level 
for $\lambda=0$, at least after broadening, while there are
none for $\lambda=1.3$, giving rise to an insulating state. 
The sharp peak at $\omega=0.0$ in the $\lambda=0.0$ case
may correspond to localized states similarly as previously reported in Ref.~\onlinecite{sen10} in the CE-states context.}
\label{Fig_dosx14}
\end{figure}

From many previous theoretical and experimental investigations, it is by now clear that the 
existence of a CMR relies on a competition between a FM metallic state 
and a charge, spin, and orbitally ordered insulating state.\cite{review,sen07,sen10} 
The new phase FM-CE 
unveiled here has all the properties needed for the insulating state, but
in addition surprisingly it has a spin FM component.
As shown in Fig.~\ref{Fig_dosx14}, the calculation of the density-of-states 
shows that this new state
has a gap at the Fermi level, thus confirming its insulating character. 
This figure shows the density-of-states
obtained from a Monte Carlo converged state as well as a for an 
idealized configuration of spins, where the latter is shown
both before and after broadening the delta-function peaks 
using a Lorentzian. In the idealized configuration, the density
of states was calculated for a fixed configuration of spins (see Fig.\ref{Fig_intx14}), 
and the lattice degrees of freedom were not taken 
into account. Hence, this calculation effectively corresponds to a zero electron-lattice 
coupling ($\lambda=0$). This is interesting in
the sense that it is known that a finite electron-lattice coupling is needed 
for the purposes of CMR. While the antiferromagnetic
coupling $J_{\rm AF}$ appears sufficient for the stabilization of the spin structure, 
the electron-lattice coupling stabilizes the 
charge-ordering characteristics. Thus, the present calculations suggest that both couplings 
play equally important roles to generate the CMR resistivity peak.

\begin{figure}[ht]
\includegraphics[clip,width=8cm,height=6cm]{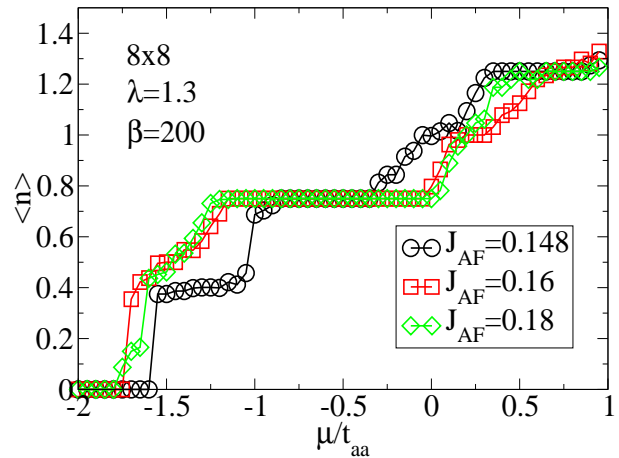}
\caption{(color online) 
Average electronic density vs. chemical potential for the 
FM state, the intermediate new state ``FM-CE'', and the 
$C_{1/4}E_{3/4}$ state, with parameters
as indicated in the figure. Note the wide plateau at $\langle n \rangle=0.75$, suggesting that the intermediate
state is quite stable, similarly as the two neighboring states FM and CE. In this calculation, $20,000$ Monte Carlo steps 
were used, where the initial $10,000$ steps are discarded for warm-up purposes.}
\label{Fig_nvsmu}
\end{figure}

The stability of these three states has been studied 
via calculations of the average carrier density
per site $\langle n \rangle$ vs. the chemical potential $\mu$. Figure~\ref{Fig_nvsmu} shows wide
plateaus at density $\langle n \rangle = 0.75$ varying $\mu$, at three different couplings representative of the
three different states, clearly indicating 
that all these states are stable in the parameter regions studied in this paper. Also, note that there is another
stable region for $\langle n \rangle=1.25$ (the $y$-axis in Fig.~\ref{Fig_nvsmu} is normalized to $\langle n \rangle = 2$,
the maximum density per site allowed by the two-orbital model), indicative of a possible particle-hole symmetry 
in the system. It is expected that CMR should also reveal itself for systems that are electron doped, but such a study 
is beyond the scope of the present publication.

The resistivity vs. temperature 
figures presented in Ref.~\onlinecite{sen10} are here reproduced for the benefit of
the reader.
These curves display changes of
over a couple of orders of magnitude near the ordering magnetic temperature, 
when parametrized as a function of the superexchange coupling
$J_{\rm AF}$. The location of the resistivity peak coincides 
with the Curie temperature $T_{\rm C}$ 
where the system presents a spontaneous first-order transition to a FM
state with a finite magnetization. This first-order transition is found 
for couplings very close to the phase boundary of the FM metallic state
(and a rapid crossover replaces the first-order transition as the distance to the
competing insulator increases).
The details of these results have been already 
discussed in Ref.~\onlinecite{sen10} and will not be
repeated here. But it is important to remark 
that the identification of the new FM-CE state in this manuscript
does not alter at all the transport CMR 
results of Ref.~\onlinecite{sen10}.

\begin{figure}[ht]
\includegraphics[clip,width=6cm,height=8cm]{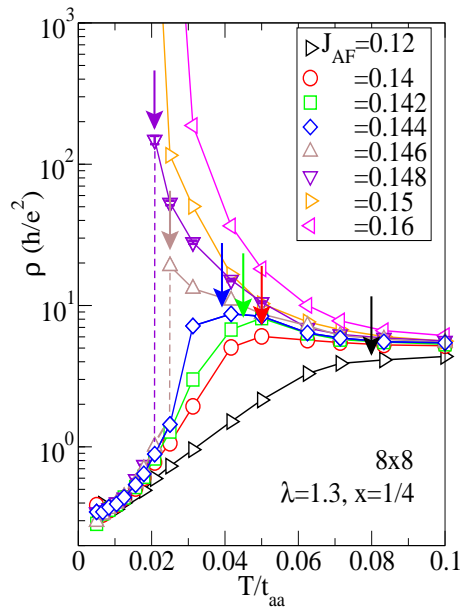}
\caption{(color online) 
Resistivity vs. temperature at $x=1/4$ doping and moderate electron-lattice coupling ($\lambda=1.3$)
parametrized as a function of the superexchange coupling $J_{\rm AF}$. A canonical CMR is observed at this doping.
The arrows indicate the Curie temperature for ferromagnetism at each value of $J_{\rm AF}$.
This result is reproduced from Ref.~\onlinecite{sen10}.} 
\label{Fig_resistivity2}
\end{figure}

\section{Spin structure factor}

In this section, the spin-structure factors of the various states discussed here are calculated 
in order to guide future neutron scattering
experiments applied to CMR manganites. 
For this purpose, here a $24\times 24$ lattice is used, and the spins are fixed  
to the various $C_{x}E_{1-x}$ patterns
as well as to the two new states discussed in the previous two sections. First, the real-space spin-spin correlations
were calculated for various spin pairs averaged over the entire lattice, $(1/N)\langle \bf{S}_{\bf i} \cdot \bf{S}_{\bf j} \rangle$. 
Then,  the spin structure factor $S({\bf k})$ was obtained by simply considering the Fourier transform of the spin-spin correlations:

\begin{equation}
S({\bf k})=\frac{1}{N}\sum_{{\bf i,j}} \langle \bf{S}_{\bf i} \cdot \bf{S}_{\bf j} \rangle
 \exp{{\it i}{\bf k}\cdot({\bf r}_{\bf i} - {\bf r}_{\bf j})},
\end{equation}

\noindent where ${\bf r}_{\bf i}$ is the lattice location of the classical spin $\bf{S}_{\bf i}$.

To better simulate the experimental situation, where preferred directions in the spin order may not be stabilized due to the formation of multiple domains, 
the states under discussion are symmetrized using various rotation and reflection operations. For example, for the simple
$E$-phase, the zigzag chains can be oriented either along the [$11$] or [$1\bar 1$] directions. The symmetry operation to transform
one into the other in this case would be a $\pi/2$-degree rotation either in the clockwise- or counter-clockwise directions.
In more complicated cases, 
such as in the newly found intermediate states, there are 
two additional states that can be obtained
by first rotating the starting state by $\pi/2$-degrees along the $z$-direction (perpendicular to the layers here investigated), 
and then taking the reflection of the resulting
state along the [$11$] or [$1\bar 1$] directions. All these states could appear in real crystals since they are all energy degenerate,
and the average $S({\bf k})$ calculated by averaging over them could be directly contrasted against the results of 
neutron scattering experiments. 

\begin{figure*}[ht]
\includegraphics[clip,width=1.6\columnwidth]{\mypath{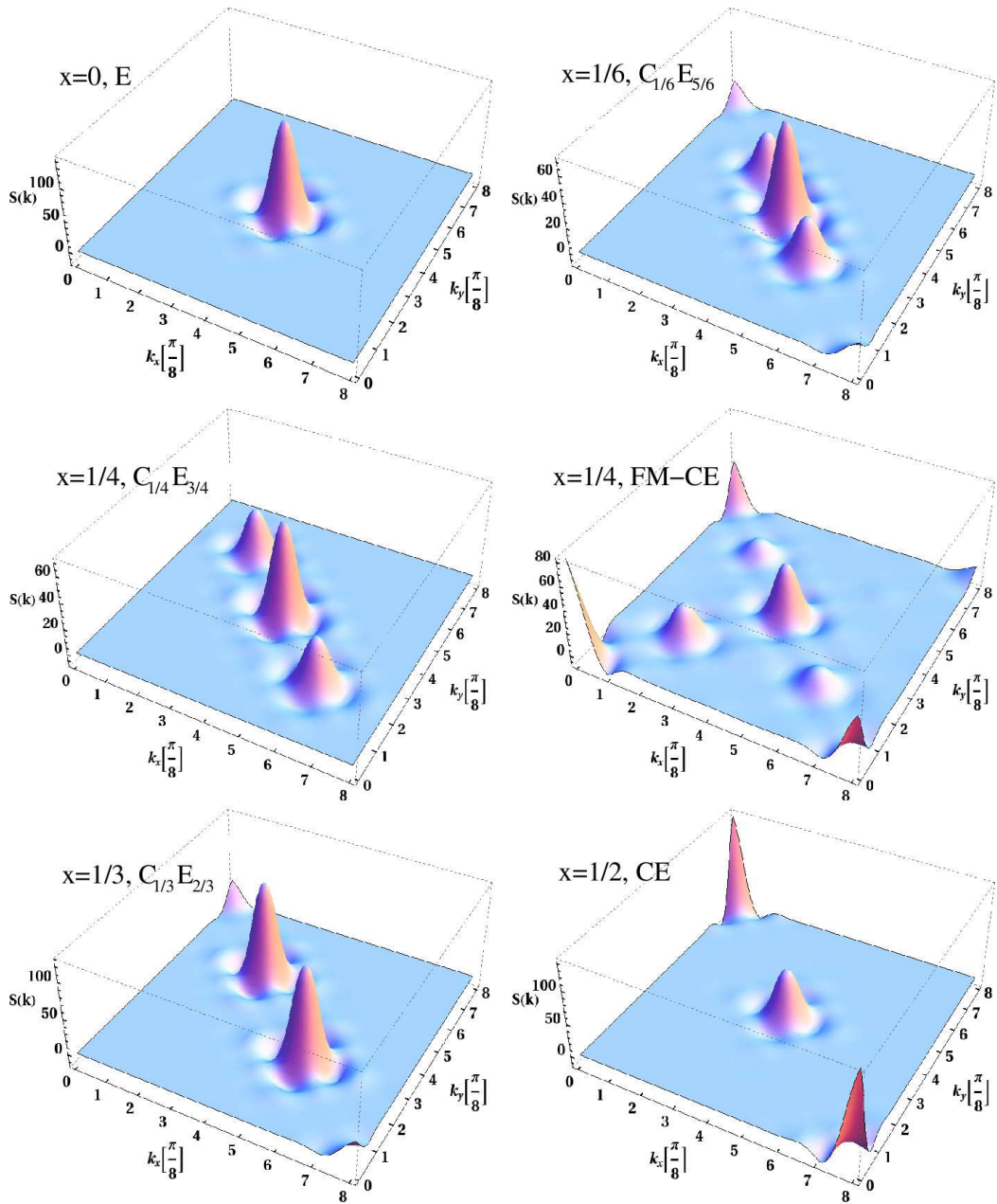}}
\caption{(color online) 
The spin structure factors for the various states at the densities indicated in the figure using a $24\times 24$ 
lattice. For each case, various rotation and reflection operations are used to symmetrize the states, as explained
in the text, for better comparison with potential neutron scattering results.}
\label{Fig_math_Sq_CE_v2}
\end{figure*}

In Figure~\ref{Fig_math_Sq_CE_v2}, the spin structure factors 
for various states are shown in the first quadrant of the Brillouin zone. 
For the previously studied $C_{x}E_{1-x}$ states, the dominant peak at 
$(\pi/2,\pi/2)$ evolves into other peaks
along the $(\pi,0)-(0,\pi)$ direction as $x$ is varied.\cite{shuhua} 
Thus, if these $C_{x}E_{1-x}$ states are
searched for with neutron scattering, 
the diagonal $(\pi,0)-(0,\pi)$ must show more intensity
than the opposite one $(0,0)-(\pi,\pi)$.

The new states FM-CE found here display interesting behavior that merits 
further discussion. These states not only
have peaks along the $(\pi,0)-(0,\pi)$ direction, induced by the CE component, 
but also along the other diagonal $(0,0)-(\pi,\pi)$ direction (although not with the same
intensity along the two diagonals). This is to be expected,
since the new states possess properties inherited from both the 
FM state and the $C_{x}E_{1-x}$ states. While the peaks
along the $(\pi,0)-(0,\pi)$ state are 
attributed to the $C_{x}E_{1-x}$ characteristics, the peaks along the
$(0,0)-(\pi,\pi)$ could similarly 
be attributed to the FM characteristics (with $(0,0)$ being the characteristic momentum of
a long-range uniform FM state). 
These results are actually similar to those reported
by Ye {\it et al.} using neutron scattering experiments applied to single-layer manganites.\cite{fye09}
Those experimental results were the first to identify 
the existence of peaks along
the $(0,0)-(\pi,\pi)$ direction in a single-layer 
manganite Pr$_{1-x}$Ca$_{1+x}$MnO$_4$, for doping levels $x < 0.5$. 
Their tentative explanation for their findings was expressed in terms of inhomogeneous electronic self-organization, where electron rich
domain walls with short-range magnetic correlations are separated from 
commensurate AF patches. 

In our present results an alternative explanation to the results of Ye {\it et al.}\cite{fye09}
can be envisioned. The notorious peak that they observed along the $(0,0)-(\pi,\pi)$ direction might
very well be due to the existence of the new state reported here
that lies in parameter space in between the FM and $C_{x}E_{1-x}$ states, in the
phase diagram varying $J_{\rm AF}$. It is clear from our $S({\bf k})$ calculations for the intermediate state that,
just like as in the neutron-scattering
results of Ye {\it et al.}, the peak located in between the ($0,0$) and ($\pi,\pi$)
peaks has a tendency to move forward in the ($\pi,\pi$) direction as the hole 
doping is increased from $x=1/4$ to $x=1/2$. While it was quite reasonable to
explain the neutron scattering results with short-range incommensurate orderings 
as was done in Ref.~\onlinecite{fye09} using
inhomogeneous states, the relevant physics
might also be explained via long-range ordering 
using the intermediate states found in the present manuscript. Further work is needed to better
clarify this important aspect of the interpretation of the neutron results.

\section{Conclusions and Summary}

In this manuscript, the region of the phase diagram 
where previous investigations~\cite{sen10} unveiled a clear CMR
effect in computer simulations has been revisited, with emphasis on the detailed properties
of the insulating state that competes with the double-exchange
induced FM metal to generate such a CMR effect. In agreement with those previous investigations,
this competing insulator is confirmed to have ``CE'' characteristics
with an associated magnetic peak in the spin structure factor $S({\bf k})$
at wave-vector ($\pi$/2,$\pi$/2) for hole density $x=1/4$. 
However, this insulating state was found to have an unexpected and far more 
complex structure, since, in addition to CE regions,
it also contains FM regions, with the combination having diagonal stripe characteristics
that alternate from CE to FM. This exotic structure 
generates extra peaks in $S({\bf k})$ due to the FM component. 
The standard $x=1/4$ CE state, without FM regions, is stabilized
upon further increasing the superexchange coupling between the $t_{2g}$ spins.
In other words, here it is reported that at $x=1/4$ and in a narrow region
of parameter space, there is a new unexpected
phase between the FM metal and the CE insulator, which is a mixture of those
two states. It appears that this new state is the competitor of the FM metallic phase
that causes the CMR effect in the Monte Carlo computer simulations. Note
that the present results are obtained in the clean limit, i.e. without
quenched disorder so this new state with mixed characteristics is intrinsic
of the standard model for manganites (although it appears in a narrow region
of parameter space). Also note that all the previous results
presented in Ref.~\onlinecite{sen10} with regards to transport and CMR effects
are unchanged and confirmed, the only modification being the characterization of the
insulating competing state that is found to be 
a mixture FM-CE as opposed to just a CE-like state. 

There are a few caveats that the reader should be alerted about with regards to
the new results presented in this paper. First, states as complex as the one
unveiled here with a FM-CE mixture in an alternating diagonal striped pattern
are difficult to analyze with regards to finite size effects. The reason is that a size
scaling analysis must be carried out only using clusters where the 
state under discussion fits properly, otherwise frustration effects will complicate
the study. The present status of computer simulations of models of manganites
prevents such an analysis to be carried out because of the rapid growth of the computer
effort as $N^4$, with $N$ the number of sites, due to the fermionic 
diagonalizations needed. Thus, the present efforts are restricted 
to two lattice size only, but it is certainly 
reassuring to find out that in both of these two
clusters the FM-CE state is stable.
Second, the study carried out here has been performed
using two-dimensional clusters, since a three-dimensional study 
with the new complex state fitting properly inside the studied lattice cannot be carried
out at present again due to the rapid growth of the effort with the number of
sites. Then, while our work in principle does apply to single-layer manganites,
it is unclear if they can be extrapolated to three dimensions assuming a
stacked arrangement of the new state. For this reason is that our predictions
for neutron scattering have {\it not} been compared against available neutron
scattering data for the three-dimensional manganites, such as LCMO, but it was contrasted
only against single-layer neutron results. Third, in this and related efforts it is
often assumed that screening effects are sufficient to suppress long-range
Coulombic forces. This is usually a reasonable assumption widely used before
in model Hamiltonians for  manganites, cuprates, and other transition metal
oxides. However, the new FM-CE state has nanoscopic regions where the electronic
densities seem to be different, i.e. the FM and CE regions do not have the same
average electronic density. Thus, another issue that must be studied in future
efforts is the stability of the present results against long-range Coulomb
repulsion.\cite{hotta-coul} 

With the caveats previously described, the results presented here suggest
a degree of complexity in the study of manganite models that is well beyond
previous expectations. The much studied CE state at $x=1/2$ with simultaneous
spin/charge/orbital order in zigzag arrangements was already considered rather
exotic in the complex oxide context. The recent theoretical efforts unveiling
highly-degenerate states involving diagonal stripes already suggested that this
simple picture may be incomplete and hinted toward
an even higher degree of complexity with similar patterns that are repeated
in a variety of ways along, e.g. one of the diagonals.\cite{SOS,shuhua} The present results
add a new layer of complexity since now the diagonal stripes alternate between
having FM and CE characteristics. This new state far from being a pathology apparently
is the key element to induce the famous CMR effect, at least in the 
computer simulations using finite clusters 
reported here and in Refs.~\onlinecite{sen07,sen10}, and in the clean limit.
Thus, a new avenue may have been opened in the study of manganites that hopefully will locate us closer
to understanding these materials and the associated CMR effect. The present effort has revealed 
a novel unexpected state that was not envisioned  before, since it emerges from
a nontrivial competition of several tendencies. 

\section{Acknowledgment}
This work was supported by the National
Science Foundation under grant DMR-11-04386.
The computational effort 
was supported in part by the 
National Science Foundation through Teragrid 
resources under grant number TG-DMR110033. 
The computations were performed on Kraken 
(a Cray XT5) at the National Institute for Computational
Sciences (\url{http://www.nics.tennessee.edu/}). 
This research used the SPF computer program and software toolkit developed at ORNL 
(\url{http://www.ornl.gov/~gz1/spf/}).

\end{document}